\documentclass[aps,prb,twocolumn,showpacs,preprintnumbers,amsmath,amssymb,floatfix]{revtex4}

\usepackage{graphicx}
\usepackage{amssymb}
\usepackage{dcolumn}
\usepackage{bm}
\begin{document}


\title{Landau Levels and Band Bending in Few-Layer Epitaxial Graphene}
\author{Hongki Min$^{1,2}$}
\email{hmin@umd.edu}
\altaffiliation[Current address: ]
{Condensed Matter Theory Center, Department of Physics, 
University of Maryland, College Park, Maryland 20742, USA}
\author{S. Adam$^{1}$} 
\author{Young Jae Song$^{1,2}$}
\author{Joseph A. Stroscio$^{1}$}
\author{M. D. Stiles$^{1}$}
\author{A. H. MacDonald$^{3}$}
\affiliation{
$^{1}$Center for Nanoscale Science and Technology, 
National Institute of Standards and Technology, 
Gaithersburg, Maryland 20899-6202, USA\\ 
$^{2}$Maryland NanoCenter, University of Maryland, 
College Park, Maryland 20742, USA\\ 
$^{3}$Department of Physics, 
University of Texas at Austin, Austin, Texas 78712, USA 
}

\date{\today}

\begin{abstract}
The carrier density distributions in few-layer-graphene systems grown
on the carbon face of silicon carbide can be altered by the
presence of a scanning tunneling microscope (STM) tip used to probe
top-layer electronic properties, and by a perpendicular magnetic field
which induces well-defined Landau levels.  Hartree approximation
calculations in the perpendicular field case show that charge tends to
rearrange between the layers so that the filling factors of most
layers are pinned at integer values.  We use our analysis to provide
insight into the role of buried layers in recent few-layer-graphene
STM studies and discuss the limitations of our model.
\end{abstract}

\maketitle

\section{Introduction} 

Progress in the preparation and isolation of highly ordered graphene
sheets over the past few years\cite{review,epireview} has led to an
explosion of interest in the properties of these two-dimensional
electron systems which are remarkably simple, yet rich in interesting
mechanical and electronic properties.  One type of graphene
system\cite{epireview} that is potentially suitable for applications
is prepared by thermal decomposition of silicon carbide (SiC).  The
unique feature of these {\em epitaxial graphene} systems is that they
tend to grow not as single layers but as few layer graphene (FLG)
systems.  The layers tend to be electrically isolated to a reasonable
degree\cite{santos,shallcross,mayou,mele,bistritzer} because of
partially controlled relative
rotations.\cite{rotatedGraphite,rotatedEpitaxial} FLG systems on SiC
can be grown as large area films that are extremely highly ordered, at
least locally, and doped by charge transfer from a carbon buffer layer,
which is a nongraphitic carbon layer between the SiC and the graphene layers.

\begin{figure}
\includegraphics[width=1\linewidth]{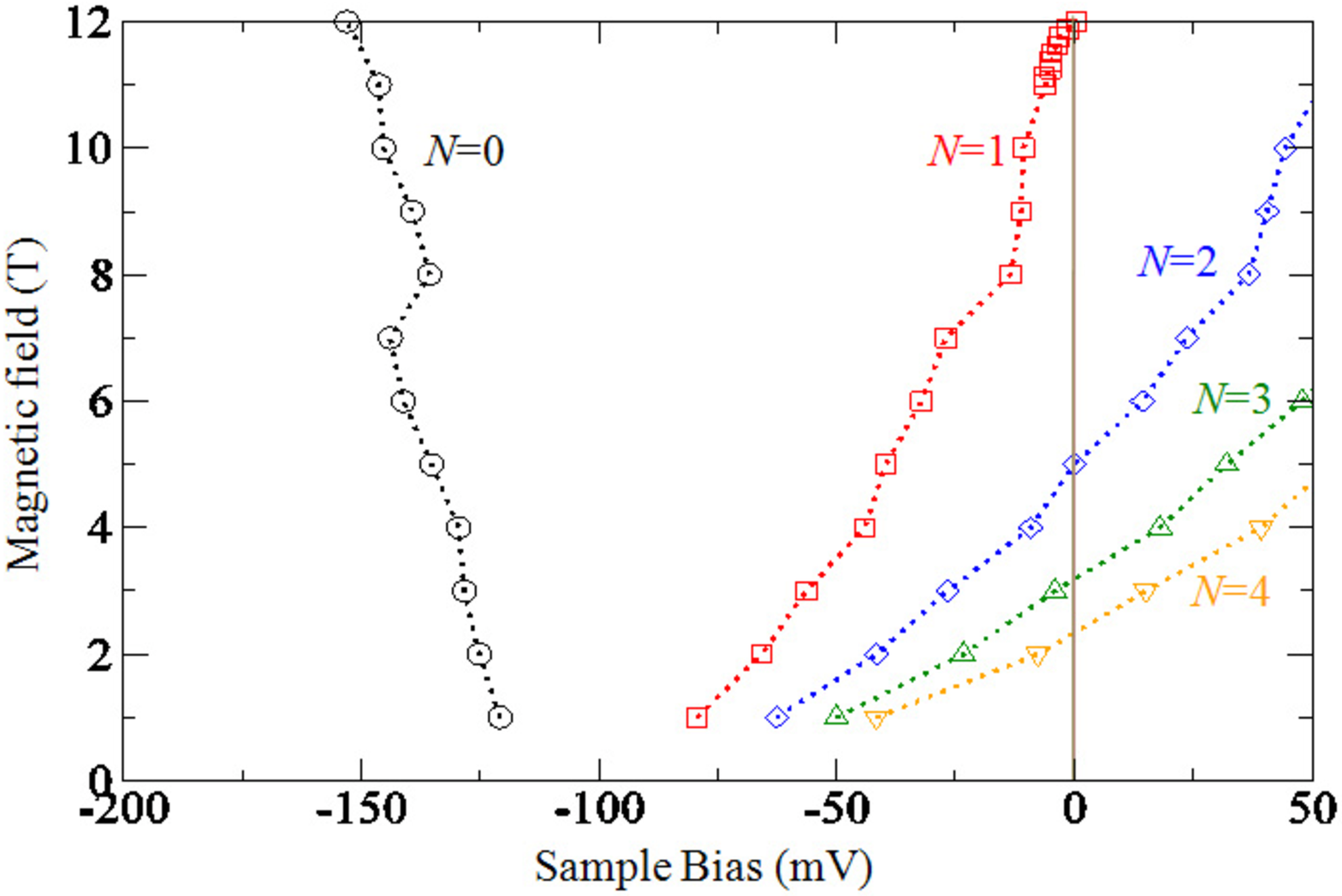}
\caption{(Color online) Landau level peak positions of epitaxial
  graphene on C-face SiC as a function of magnetic field obtained from
  the STM measurements in Ref.~\onlinecite{Joe2}. The position of
  each Landau level is averaged over spin and valley split
  $dI/dV$ peaks when these can be separately resolved. \cite{note_errors} 
  $dI/dV$ peak positions at   finite tip-sample bias can be influenced 
  by tip-sample band bending and by redistributions 
  of charge in the FLG system as explained in detail below.}
\label{fig:LL_STM}
\end{figure}

This paper addresses the distribution of charge carriers across the FLG system.  
Recent measurements of Landau level spectra\cite{rotated_LL,Joe1} and angle-resolved photoemission\cite{rotated_arpes} 
for FLG systems grown on the carbon face of SiC show characteristics of decoupled monolayer graphene 
rather than coupled graphene multilayers. This behavior is likely due to the relative rotations between the layers.
Our work is motivated in part by an interest in
understanding the confusing\cite{reftransport} transport properties
of these systems, which must be strongly dependent on carrier charge
distribution across the weakly coupled layers.  Our immediate
motivation, however, is provided by recent\cite{Joe1,Joe2} scanning
tunneling microscopy (STM) Landau level spectroscopy studies of FLG in
the presence of an external magnetic field.  Traces of the Landau level
positions can be extracted from such spectra as is illustrated   
in Fig.~\ref{fig:LL_STM}.  Although STM directly
probes electronic properties in the top layer of a FLG system, there
is evidence that top layer properties can be altered, sometimes
qualitatively, by correlations with electrons in submerged layers.  
 
If the density in the top layer were fixed, the Fermi level would be
pinned to one of the Landau level energies except at the discrete
field strengths which yield integer filling factors 
when the Fermi level is in between Landau levels.  In practice,
experiment shows the opposite behavior.  The Landau levels tend to be
pinned away from the Fermi energy, an effect that is particularly
striking in the field range between 8~T and 10~T.  At higher fields,
the Landau levels split through valley and spin splitting so that even
in the field range near 12~T, the split Landau levels avoid the Fermi
energy.

The most dramatic effect seen in these experiments is splitting within
spin- and valley-split $N$=1 peaks in the density of states (DOS) as they pass through the Fermi
level.\cite{Joe2} This peculiar, fractionally filled Landau level
gives evidence for a correlated-electron state that is stable when the
$N=1$ Landau level of the top layer is half-filled.

While the precise nature of this state remains mysterious, its
formation might depend only on correlations among top layer electrons;
however, if one of the submerged layers is also partially filled under
the same tip-biasing and field conditions, then this
fractionally filled Landau level could depend essentially on
correlations between electrons in different layers. Since half-filling
does not favor the formation of especially stable states in an
isolated layer, the latter possibility appears likely.  In the
strong-magnetic-field quantum Hall regime with fully formed Landau
levels only weakly broadened by disorder, correlations are strongest
when Landau levels are partially filled.  One goal of the model
developed in this paper attempts to provide a basis for estimating
which layers contain partially filled Landau levels as the magnetic
field strength varies.

Our paper is organized as follows. In Sec. II, we explain our model
for carrier distribution in a few layer graphene system in which the
buffer layer acts as a reservoir for carriers.  We assume that the
growth-dependent buffer layer properties determine the position of the
Fermi energy relative to the Dirac point of the bottom graphene layer.
Electron-electron interactions are included only at the Hartree level.
At zero magnetic field carriers reside mainly in the layers closest to
the buffer and the density-of-states in the top layer is small.  When
a perpendicular magnetic field is applied, the Fermi level tends to be
pinned near one of the filling factors ($\nu=\pm 2,\pm
6,\pm10,\ldots$) at which the integer quantum Hall effect occurs in a
graphene layer, implying that charge must be transferred between
layers as a function of field.  In Sec. III, we discuss how an STM
tip can be included in such a model.  When an STM tip is introduced to
study the electronic properties of the top layer, its carrier density
tends to be altered with a sign and magnitude that is
strongly dependent on the tip work function.  The STM studies in
Refs.~\onlinecite{Joe1,Joe2} show that the top-layer is $n$-type for
the tip used in those experiments, so that carrier densities peak not
only near the buffer layer but also near the top-layer. In Sec. IV,
we use this basic theoretical picture to develop a theory of STM
Landau level spectroscopy in FLG, comparing where possible with STM
data. We find that as the sample-tip bias and the magnetic field are
varied, charge tends to rearrange to achieve integer filling factors
in as many of the FLG layers as possible.  In Sec. V, we conclude
with a brief summary and some suggestions for future experimental and
theoretical work.

\section{Few-Layer-Graphene Model}

\begin{figure}
\includegraphics[width=0.8\linewidth]{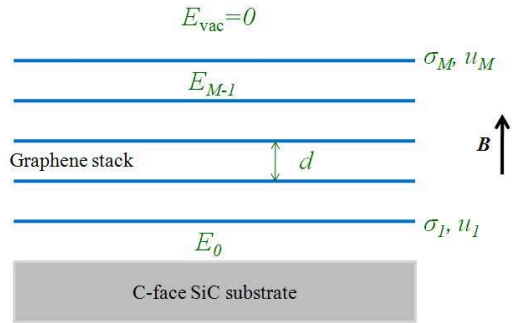}
\caption{(Color online) Schematic illustration of a
  few-layer-graphene system without an STM tip.
The buffer layer between the SiC and the graphene layers was omitted for simplicity.} 
\label{fig:model_no_tip}
\end{figure}

We estimate carrier charge distribution in a FLG system grown on carbon-face SiC substrates
using the model summarized schematically in Fig.~\ref{fig:model_no_tip}.
Earlier work considered the charge distribution on mono and bilayer
graphene\cite{Kopylov2010} and for multilayer graphene\cite{Datta2009} in the
continuum limit, both in the absence of a magnetic
field, our main interest.  In Fig.~\ref{fig:model_no_tip}, the
graphene layers are labeled by integer numbers starting from label $1$
for the layer closest to the buffer layer to $M$ for the top layer.
(The buffer layer between the SiC and the graphene layers, which acts as a reservoir for carriers, was omitted for simplicity.)
In equilibrium all layers share the same chemical potential $\mu$. The
Dirac point in layer $i$ is shifted by its local electric potential $u_{i}$. 
The energy spectrum of each layer is that of monolayer graphene with the Dirac point shifted by $u_i$.
It follows that the charge density in layer $i$, $\sigma_i$,
satisfies
\begin{equation} 
\sigma_i=\frac{{\rm sgn}(\mu-u_i)}{
  \pi}\left(\frac{\mu-u_i }{ \hbar v}\right)^2. 
\end{equation}
The potential energy $u_{i}$ is in turn evaluated from the
charge densities using the Poisson equation which implies that the
electric field $E_{i}$ between layer $i$ and layer $i+1$ satisfies
\begin{equation}
\label{poisson} 
\epsilon \, (E_{i}-E_{i-1}) = 4 \pi (-e) \sigma_{i}.
\end{equation} 
The dielectric constant $\epsilon$ in Eq.~(\ref{poisson}) accounts for
the polarizability between graphene sheets.  Here we choose
$\epsilon=1$; we have found that changing the value of $\epsilon$ does
not qualitatively alter the main results of this paper.

Since the electric field in the vacuum above the top ($M$-th) layer
$E_{\rm vac}$ must vanish 
in the absence of an STM tip,
the electric fields between all graphene sheets
are readily evaluated iteratively given the charge densities
$\sigma_{i}$.  Starting from layer $1$ and adding a contribution due
to the electric field between a layer and the layer above gives:
\begin{equation} 
u_{i+1} = u_{i} + e d E_{i},
\end{equation} 
where $d=0.335$ nm is the interlayer separation between graphene
layers.  Our neglect of interaction effects beyond electrostatics is
supported in the zero magnetic field limit by recent Green's function 
screened Coulomb (GW) many-body calculations\cite{Profumo} by Profumo 
{\em et al}. As we discuss below, exchange and correlation effects are 
likely to be more important in large magnetic fields.

We model the role of the buffer layer by assuming that its
equilibration with the bottom graphene layer fixes the value of
$\mu-u_{1}$.  It is known that the carrier density of the graphene
system is sensitive to the microstructure of the disordered buffer
layer, and hence to FLG growth conditions.  For a given sample, some
carriers remain after the bonding between the buffer layer and the SiC
substrate is established.  Energy in the system is lowered as these
electrons are transferred to the $\pi$-bands of the first graphene
layer.  $u_{1}$ is determined by a balance between the chemical
driving force for the transfer and the band and electrostatic energy
cost of adding electrons to the graphene.  In assuming that $u_{1}$ is
independent of field, as we do below, we are taking advantage of
the fact that the Landau-level separation at the Fermi energy of the
first graphene sheet is small compared to $\mu-u_{1}$. In modeling STM
data on a particular sample, imperfect knowledge of the most
appropriate value for $\mu-u_{1}$ is an important source of
uncertainty that limits predictive power. 
For the calculations described below we choose $\mu-u_{1}=360$ meV, an
estimate that is motivated by spectroscopic measurements
\cite{fermi_energy} in multilayer graphene grown on the C-face of the SiC
substrate.  In the rest of this paper
we choose our zero of energy so that $u_{1}=0$.

\begin{figure}
\includegraphics[width=1\linewidth]{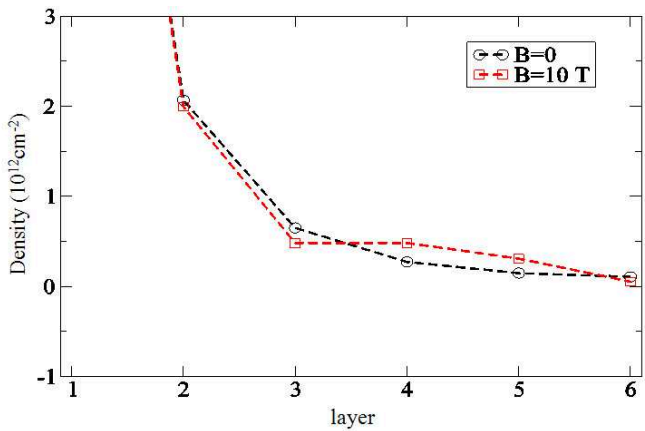}
\caption{(Color online) 
 Charge density vs. layer index for a 6-layer FLG system
 at $B=0$ T and $B=10$ T. 
 Here chemical potential $\mu=360$ meV and temperature $T=30$ K were used.}
\label{fig:layer_charge_no_tip}
\end{figure}

To explain how band and electrostatic energies combine to determine
carrier distributions, we first consider double-layer graphene with a
chemical potential $\mu>0$.  From the Poisson equation with
$E_{\rm vac}=0$, the electric field between layers, $E$, satisfies
$\epsilon E= 4\pi e \sigma_2$. The potential energy of the top layer
is $u_2=eEd$.  It follows that
\begin{eqnarray}
\sigma_1&=&\frac{1}{\pi}\left(\frac{\mu}{ \hbar v}\right)^2, \\
\sigma_2&=&\frac{sgn(\mu-u_2)}{ \pi}\left(\frac{\mu-u_2 }{ \hbar v}\right)^2 \nonumber \\
&=&\frac{sgn(\mu-u_2)}{ \pi}\left(\frac{\mu}{ \hbar v}\right)^2 f(\beta) \nonumber
\end{eqnarray}
where
\begin{equation}
f(\beta)=(\sqrt{\beta^2+2\beta}-\beta)^2
\end{equation}
and $\beta={\epsilon (\hbar v)^2 / (8 e^2 d \mu)}$ is a unitless
quantity which controls the energy balance of charge moving between
layers.  Here $v$ is the $\pi$-band velocity at the Dirac point which
is proportional to the intralayer hopping energy $\gamma_0$.
For $\mu=360$ meV, $\epsilon=1$, and $\gamma_0=3$ eV,
$\beta\approx0.29$ and $f(\beta)\approx0.28$; thus the top layer has
28\% of the bottom layer charge.

For multilayers with more than two layers, the distribution follows
from a simple numerical calculation. 
The layer charge density is calculated by integrating the Landau level 
density of states weighted by the Fermi factor for the appropriate chemical potential. 
Then from the resulting layer densities, layer potentials are calculated using the Poisson equation 
in Eq.~(\ref{poisson}). This process is repeated until a self-consistency is reached.

The charge distribution for
a decoupled 6-layer graphene stack at $B=0$ T and $B=10$ T calculated
with the same parameters used in the double layer graphene, is shown
in Fig.~\ref{fig:layer_charge_no_tip}. For $B=0$ T, layers above the
bottom layer have in total 32\% of the bottom layer charge and this
ratio is almost independent of the number of layers. As the magnetic
field is turned on, the charge distribution is altered due to the
formation of Landau levels, particularly in the low-density layers
with a Fermi level near the Dirac point.  Because a Landau level
appears precisely at the Dirac point in a graphene sheet, a magnetic
field causes a peak in the density-of-states to appear at the same
energy at which the density-of-states vanishes in the absence of a
magnetic field. This feature of graphene physics strengthens
magnetoelectric effects associated with Landau level quantization. 

\section{STM Tip Model} 

\begin{figure}
\includegraphics[width=1\linewidth]{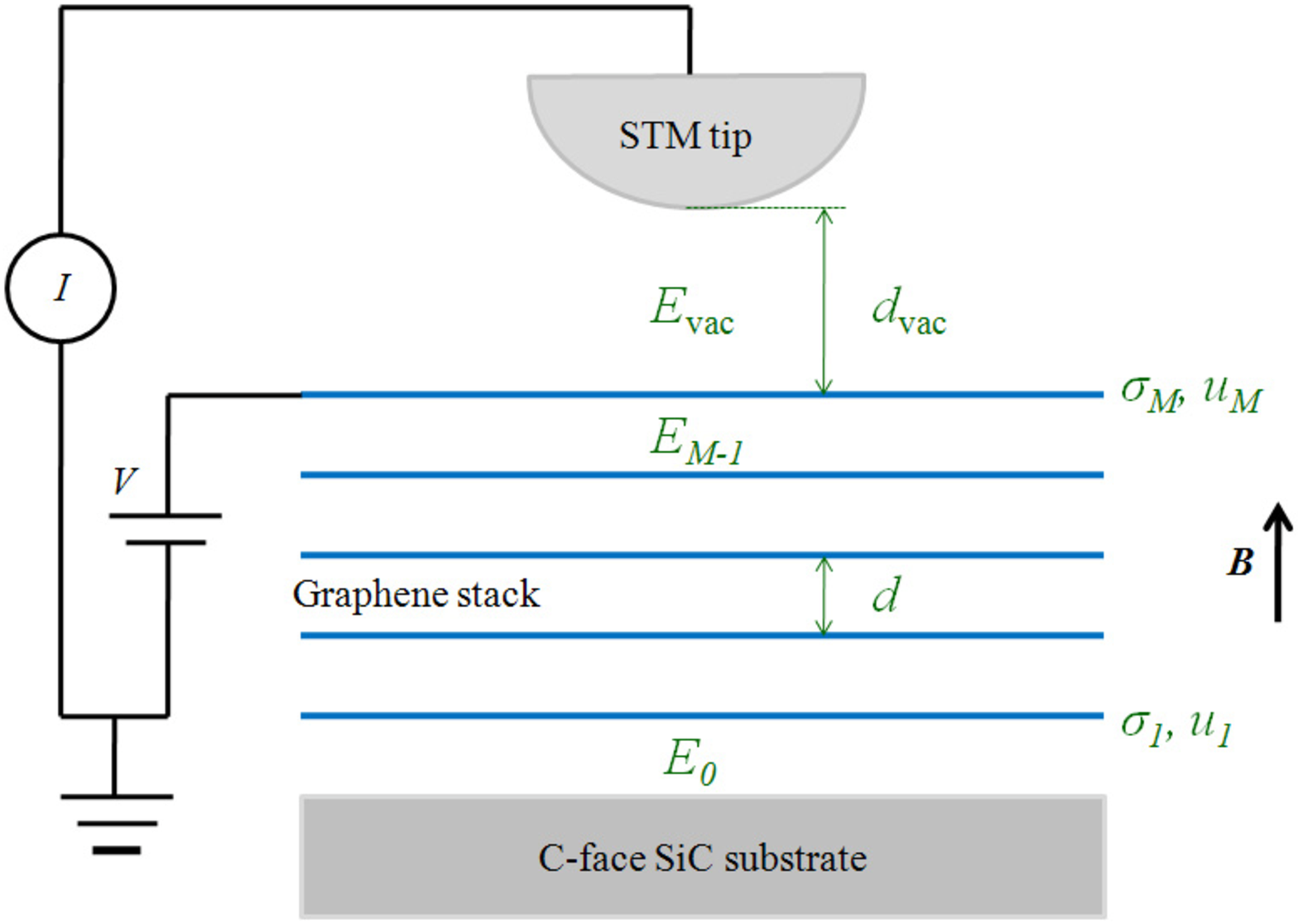}
\includegraphics[width=1\linewidth]{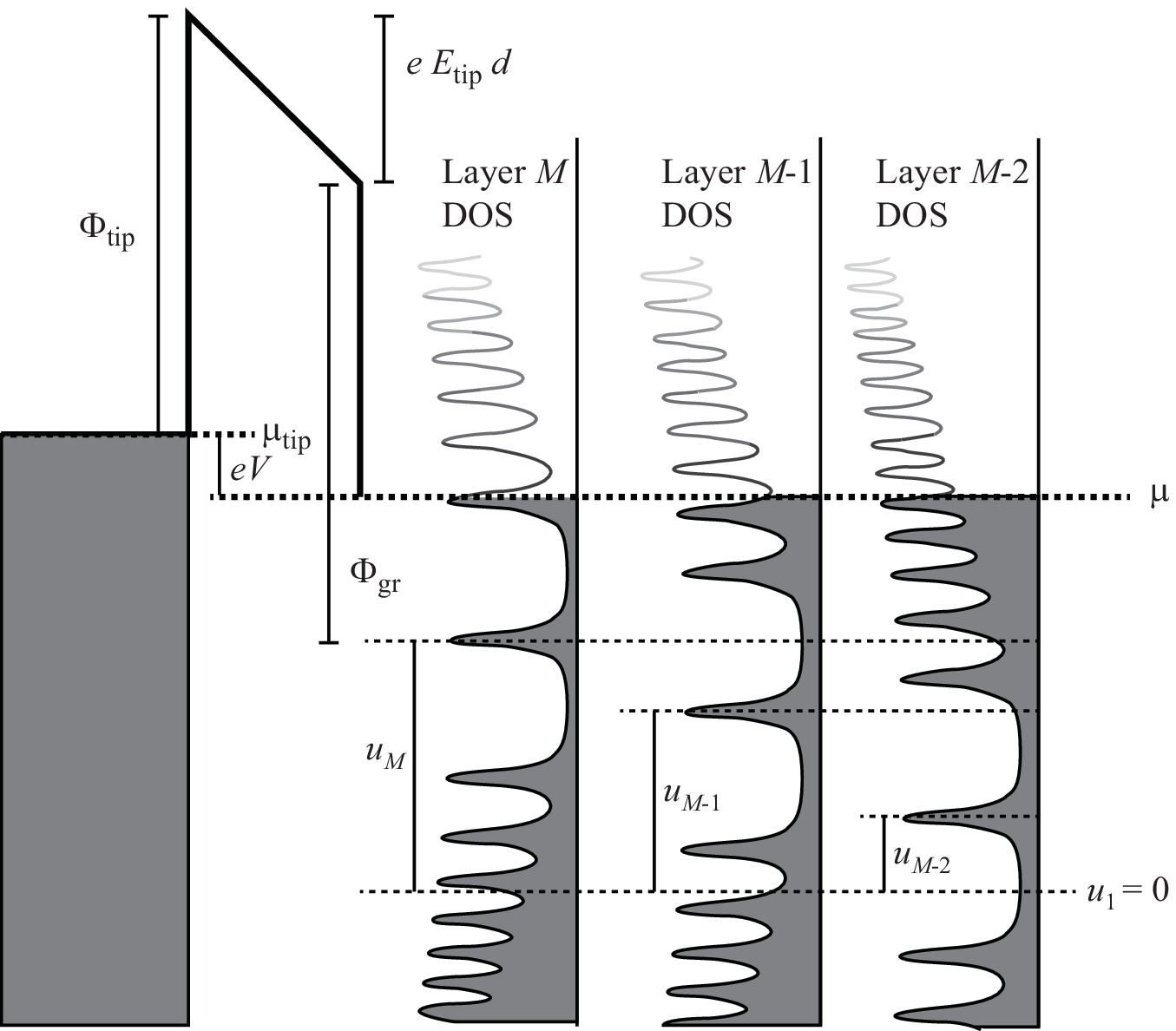}
\caption{(Color online) Upper panel shows schematic illustration of a
  few-layer-graphene system with an STM tip.  The buffer layer between the SiC and the graphene layers was omitted for simplicity.
Lower panel is the  energy level diagram for this system.} 
\label{fig:model_tip}
\end{figure}

Figure \ref{fig:model_tip} shows a schematic illustration of a FLG
system with an STM tip. We model an STM tip as an additional layer
which acts as a top gate electrode. The distance between the tip and
graphene surface is taken as $d_{\rm vac}=1$ nm.\cite{joe_book}

Experimentally, it is found that the graphene work function, i.e., the
energy to take an electron from the Fermi energy to vacuum, depends on
the charge on the surface layer.  However, it is also found that the
energy to take a graphene electron from the Dirac point to vacuum does not
change as a function of the charge density.\cite{work_function}  We denote 
the latter as $\Phi_{\rm
  gr}$.  In general, the work function of the tip ($\Phi_{\rm tip}$)
and the graphene layer [$\Phi_{\rm gr}-(\mu-u_{M})$] are different.  This
difference in work functions, $\Phi_{\rm gr}-(\mu-u_{M})-\Phi_{\rm
  tip}\equiv\Phi-(\mu-u_{M})$ leads to charge transfer between the surfaces
when they are electrically connected and induces an electric field
between the surface and tip. As seen in the bottom panel of
Fig.~\ref{fig:model_tip}, the electric field satisfies
\begin{equation}
\mu+eV=\mu^{\rm tip}=u_M+e E_{\rm vac} d_{\rm vac}+\Phi
\label{eq:evac}
\end{equation}
as a voltage $V$ is applied between tip and sample.  

\begin{figure}
\includegraphics[width=1\linewidth]{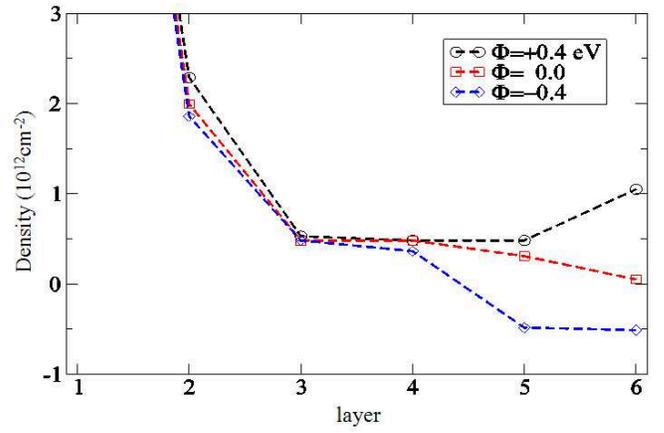}
\caption{(Color online) Charge density in the layers of a six layer
  stack at $B=10$ T for $\Phi=+0.4$ eV, $0$ eV, and $-0.4$ eV when the
  tip-sample bias is zero.  These curves were obtained using $\mu=360$
  meV and $T=30$ K. By construction, electrons have a positive 
carrier density, while holes have negative density (see text for details).}
\label{fig:layer_charge_tip}
\end{figure}

In STM spectroscopy a new tunneling transport channel opens up, giving
rise to a $dI/dV$ peak whenever the chemical potential of the STM tip
is aligned with one of the top-layer Landau levels.  The experimental
$dI/dV$ peaks therefore identify the tip-sample bias voltages at which
the following resonant tunneling conditions are satisfied:
\begin{eqnarray} 
\label{eq:dIdV_peak}
\mu+eV&=&u_{M}+\varepsilon_{N}(B),
\end{eqnarray} 
where $\varepsilon_{N}(B)=sgn(N)\sqrt{2|N|e\hbar v^2 B/c}$ is the
graphene sheet Landau level energy.  To illustrate the effect of the
tip, in Fig.~\ref{fig:layer_charge_tip} we calculate the charge
distribution of each layer at $B=10$ T for $\Phi=+0.4$ eV, $0$ eV, and
$-0.4$ eV when the tip-sample bias $V$ is zero.

For $\Phi=0$ eV and $V=0$ V,
the FLG charge
distribution is identical to the distribution without an STM tip at
$B=10$ T shown in Fig.~\ref{fig:layer_charge_no_tip}.  For non-zero $\Phi$,
however, an electric field between the tip and sample surface is
induced and distorts the layer charge distribution even at zero
tip-sample bias.

\section{FLG Landau-Level Tunneling Spectroscopy}

At weak fields, $u_{M}$ and $E_{\rm vac}$ are approximately constant
so that the spacing in electronvolts between $dI/dV$ peaks matches
the energetic separation between top-layer Landau levels.  The
spectroscopy data can therefore be used to measure the Dirac velocity
parameter which characterizes the energy scale of the graphene layer's
Dirac cones. In the strong-field limit, however, the density-of-states
in each graphene layer is altered, and this in turn alters the
densities at which equilibria are established between adjacent layers.
It follows that both $u_{M}$ and $E_{\rm vac}$ depend on field.  One
goal of our calculations is to estimate the magnitude and character of
this effect.

\begin{figure}
\includegraphics[width=1\linewidth]{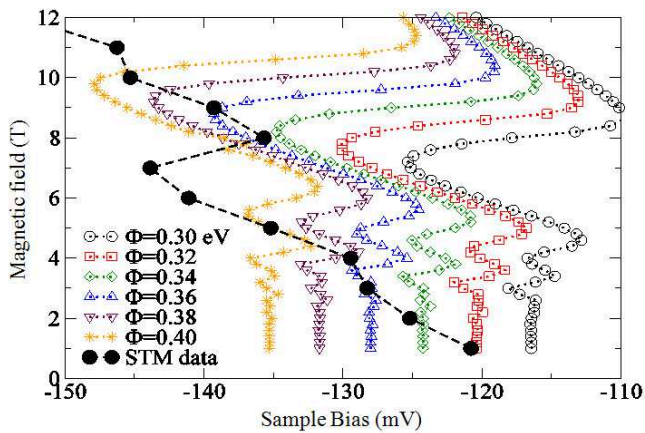}
\caption{(Color online) The field dependence of the $N=0$ Dirac-point
  Landau level for different workfunction parameters $\Phi$.\cite{note_errors} 
  Here $\mu=360$ meV and $T=30$ K were used.}
\label{fig:LL_th_n0}
\end{figure}

To illustrate this effect we first examine the field-dependence of the
$N=0$ Dirac-point Landau level feature, plotted in
Fig.~\ref{fig:LL_th_n0}.  Since $\varepsilon_{N=0}(B)\equiv 0$, the
field-dependence of this spectral feature is due entirely to the
field-dependence of charge distributions in the FLG system. The
calculations in Fig.~\ref{fig:LL_th_n0} were carried out at a finite
temperature $T=30$ K, in part to crudely model the Landau-level
smearing influence of disorder.  At weak fields the $N=0$ $dI/dV$
peak's position is independent of field as expected.  The position of
these peaks is primarily dependent on the model's workfunction parameter $\Phi$.
  
As illustrated in Fig.~\ref{fig:LL_th_n0}, the position of the
experimental peak at $eV \approx -135$ meV and at $B=5$ T is reproduced approximately by
setting $\mu=360$ meV, which leads to $u_{M} \approx 225$ meV and
$\Phi \approx 400$ meV. The slow downward drift in the weak field
$N=0$ Dirac point peak with increasing magnetic field is not
reproduced by our calculation, and could be due to an increase in the
strength of exchange and correlation effects in FLG with magnetic
field. The strong variations in peak positions with field that begin
at around $6$ T are the quantizing-magnetic-field effects on which we
will focus in the remainder of this paper.

\begin{figure}
\includegraphics[width=1\linewidth]{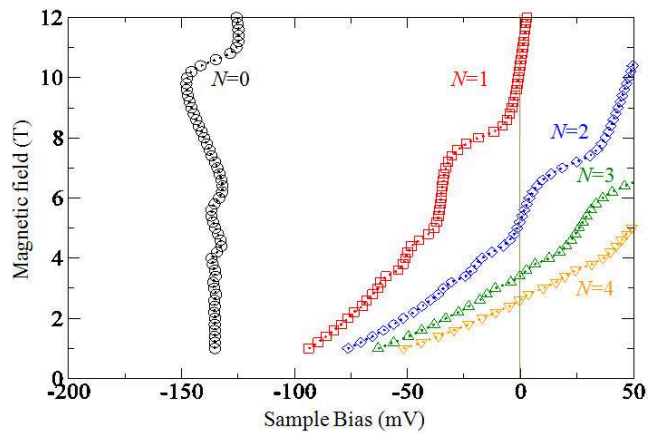}
\caption{(Color online) Theoretical prediction of top layer Landau-level peak
  positions as a function of magnetic field for $\Phi=400$ meV and
  $\mu=360$ meV at $T=30$ K.}
\label{fig:LL_th}
\end{figure}

The influence of Landau quantization on $dI/dV$ spectra is illustrated
in more detail in Fig.~\ref{fig:LL_th}, which shows the
prediction of the theoretical model for $\Phi=400$ meV and $\mu=360$
meV at $T=30$ K.

When the Landau level energies in a particular layer are far away from
the Fermi energy, the layer filling factor $\nu_i= 2\pi \ell^2
\sigma_i$, where $\ell=\sqrt{\hbar c / e B}$ is pinned at one of the
full-Landau-level filling factor values: $\nu_i=\pm 2, \pm 6, \pm 10,
\cdots$.  For a fixed filling factor the carrier density in a layer
increases with field and its Landau level energies therefore increase
due to electrostatic repulsion.  The increase in density must be
achieved by charge transfer from other layers.  When a Landau level
in a layer is close to the Fermi level, the density in that layer will
tend to decrease as its Landau level empties with increasing field.
This is the source of charge transferred to other layers.  This
behavior contrasts with that of an isolated system with fixed charge
density in which integer filling factors occur only at isolated field
values and successive Landau-level energies are pinned to the Fermi
level.
 
\begin{figure}
\includegraphics[width=1\linewidth,height=2in]{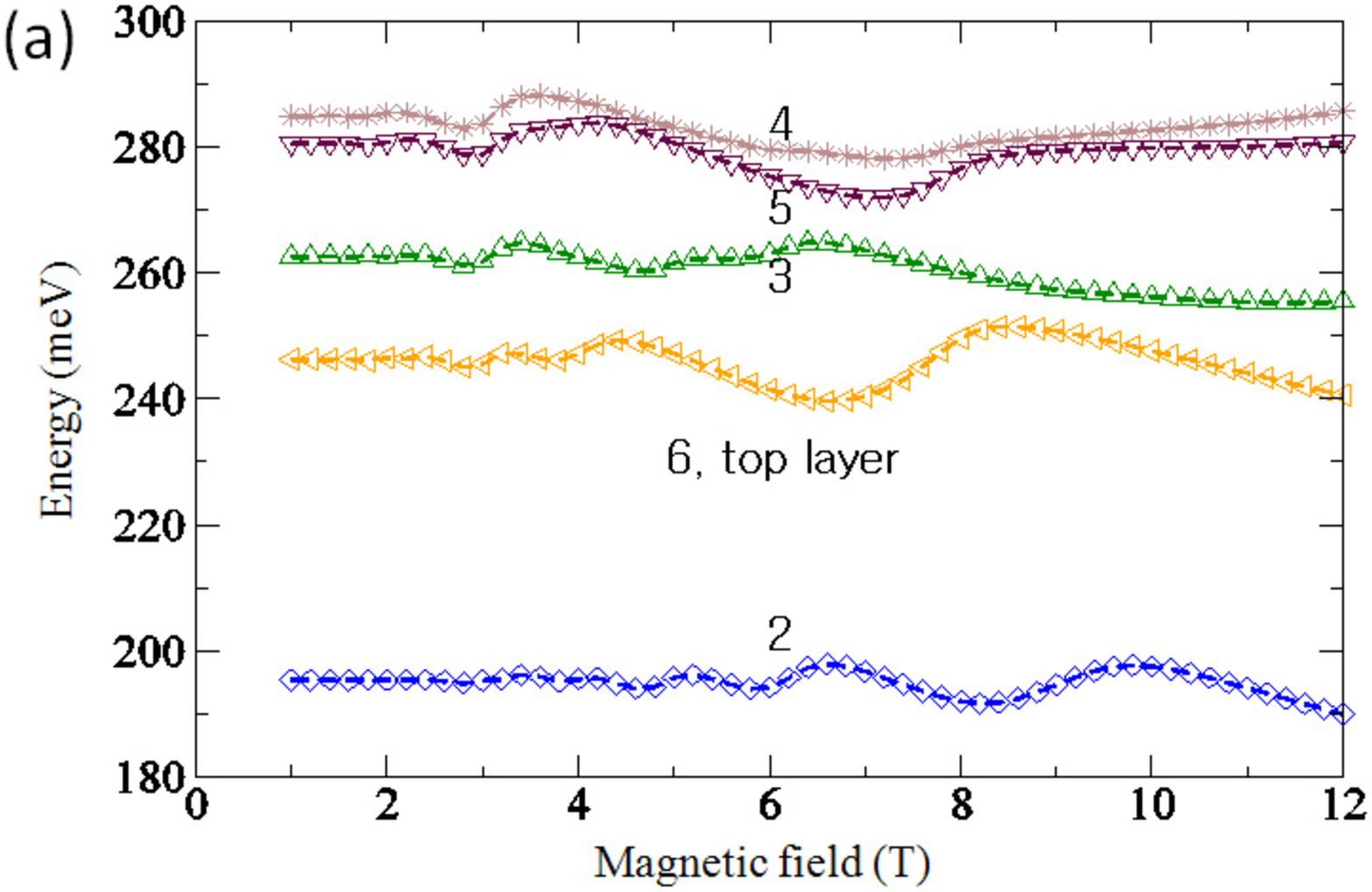}
\includegraphics[width=1\linewidth,height=2in]{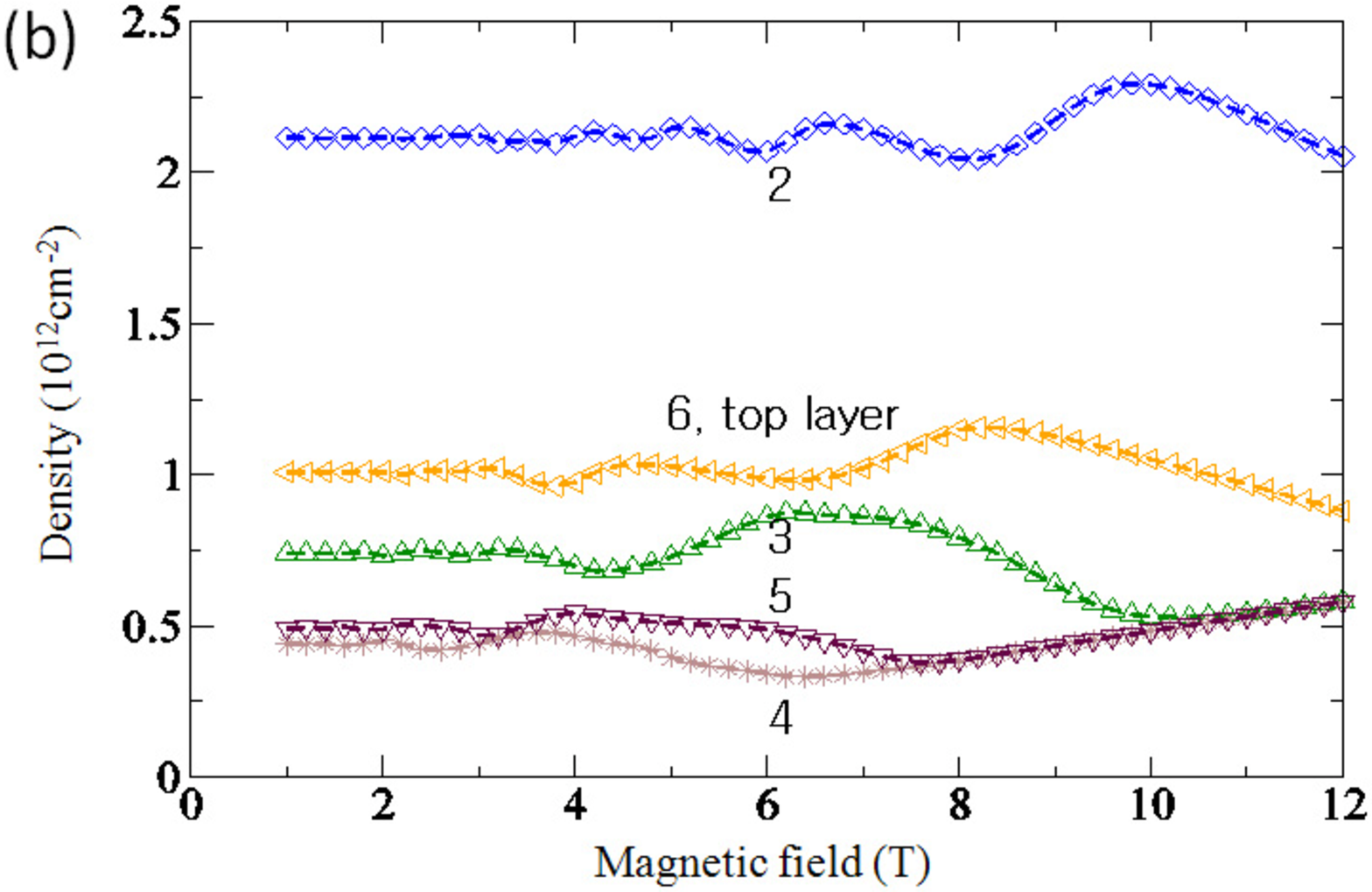}
\includegraphics[width=1\linewidth,height=2in]{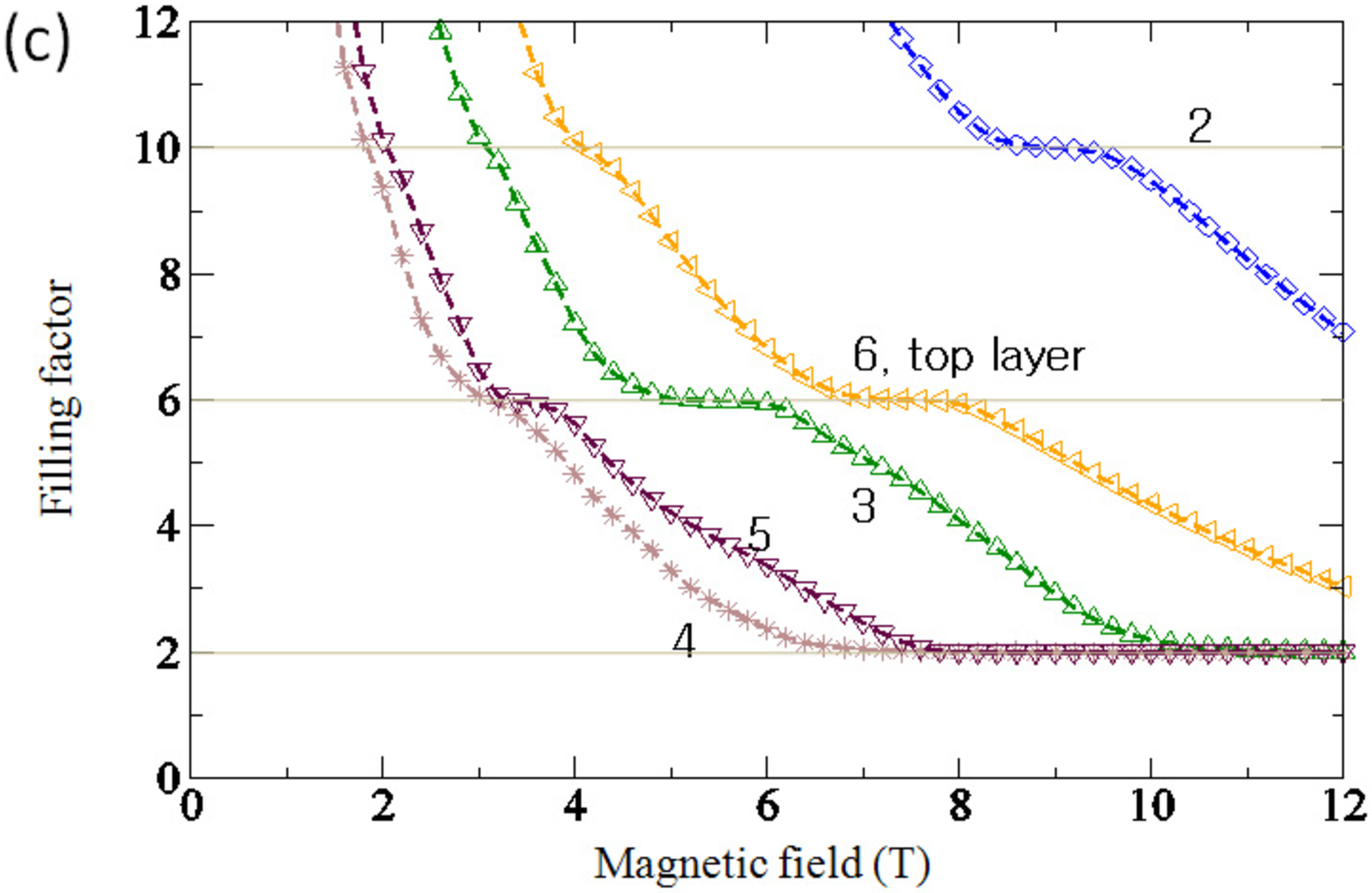}
\caption{(Color online) Electrostatic energy, density and filling
  factor of each layer as a function of magnetic field for $\Phi=400$
  meV, $\mu=360$ meV and $T=30$ K at zero tip-sample bias voltage.
  The filling factor of the bottom layer (not shown) exceeds $\nu=12$
  over the field range considered.  
  Numbers indicate layer numbers, with 6 being the topmost layer.} 
\label{fig:layer_field}
\end{figure}
  
We refer to layers which have partially filled Landau levels as {\em
  active} and to layers which have full Landau levels as {\em
  inactive}.  Since the total filling factor is a smooth function of
field, at least one layer must be active at generic field values.
Strong interlayer correlations are likely when two or more layers are
active.  It would be surprising if interlayer correlation effects were
not important, given the relationship between the important length
scales in the problem.  At 10~T, the total graphene layer thickness $M
d\approx 2$ nm is typically much less than the average separation of
electrons within one layer $1/\sqrt{\sigma_i}\approx 10$ nm as well as
the magnetic length $\ell=\sqrt{\hbar c / e B}\approx 8.1$ nm.  Such
correlations could give rise to a state with spontaneous interlayer
coherence\cite{Joe2} among other possibilities.  As shown below, our
model calculations provide estimates of the field ranges at which two
or more layers become active.

For the parameters 
of Fig.~\ref{fig:LL_th} the model predicts 
that the top layer is active between $B=5$ T and $B=7$ T.  In
this field range the $N=2$ Landau level is pinned to the Fermi level
and the filling factor varies between $\nu=10$ and $\nu=6$.\cite{note1}  
The top layer is then briefly inactive before becoming active again
above 8 T when the $N=1$ Landau level is pinned to the Fermi level.
In inactive field ranges, the density in the top layer is proportional
to magnetic field and the energies of all levels in that layer
increase.  In the active regions, the density tends to decrease and
the rate at which energy levels increase with field for $N > 0$ is
suppressed by the decrease in density.
This is only a tendency, however,
since the electrostatic energy in the top layer depends on the
densities at all layers. The evolution of the STM spectrum also
depends on the evolution of charge density in the submerged layers
that are not directly probed by the STM.  Note that the distribution
of charge among the FLG layers also depends somewhat on the tip-sample
bias voltage.  Figure \ref{fig:layer_field} shows the electrostatic
energy, density, and filling factor of each layer as a function of
magnetic field at zero bias voltage. 
These results are consistent with the preceding discussion.

\section{Discussion and Conclusions} 

\begin{figure}
\includegraphics[width=1\linewidth]{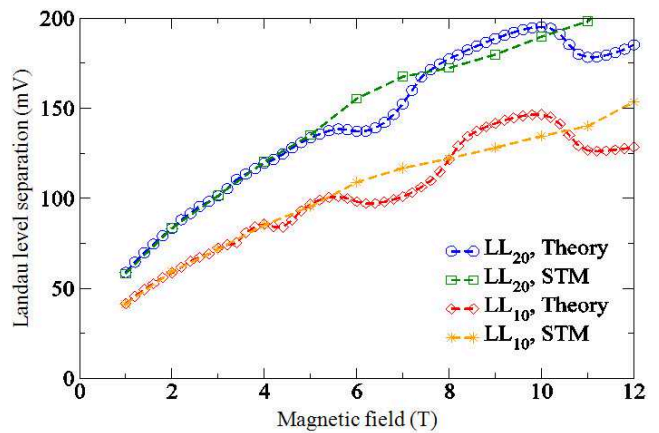}
\caption{(Color online) Landau level separations between $N=0$ and
  the $N=1$ ($N=2$), denoted as ${\rm LL}_{10}$ (${\rm LL}_{20}$) for
  theoretical calculations and STM measurements.\cite{note_errors} 
  For the theoretical calculations, $\Phi=400$ meV and $\mu=360$ meV 
  at $T=30$ K were used.}
\label{fig:LL_separation}
\end{figure}

Our theoretical model does not account for exchange and correlation
effects, which can alter the energy change associated with adding
electrons to empty (or partially filled) Landau levels and the energy
change associated with removing electrons from full (or partially
full) Landau levels.  Systematic discrepancies between present theory
and experiment likely signal these neglected interaction effects.
These discrepancies include the low field variation of the zeroth
Landau level energy (Fig.~\ref{fig:LL_th_n0}) and the pinning of the first Landau level away
from the Fermi energy rather than at it (Fig.~\ref{fig:LL_STM}).

In Fig.~\ref{fig:LL_separation}, for example, we compare experimental
and theoretical energy separations between $N=0$ and the $N=1, 2$
$dI/dV$ features as a function of magnetic field strength.  Even at
the Hartree level there are additions to the $\sqrt{B}$ band energy
contribution due to changes in electrostatic energies with tip-sample
bias voltages indicating overestimation of the electrostatic tip-gating effects.
 
We note that the filling-factor dependent features in the
field-dependence of the Landau level energies [Fig.~\ref{fig:layer_field}(c)] are weaker in experiment than in
this Hartree theory.  We believe that these differences mainly reflect
exchange and correlation energies which mitigate electrostatic
effects.  The presence of strong correlation effects in this field
range is apparent in the experimental interaction-induced Landau level
splittings which have been suppressed in Fig.~\ref{fig:LL_separation}
by averaging over all experimental features identified with $N$=0 and $N$=1.

The approximately linear reduction with field (at weak fields) 
of the bias voltage at which the $N=0$ peak is observed (see 
Fig.~\ref{fig:LL_th_n0}) is completely absent in theory and unexplained
at present.

\begin{figure}
\includegraphics[width=1\linewidth]{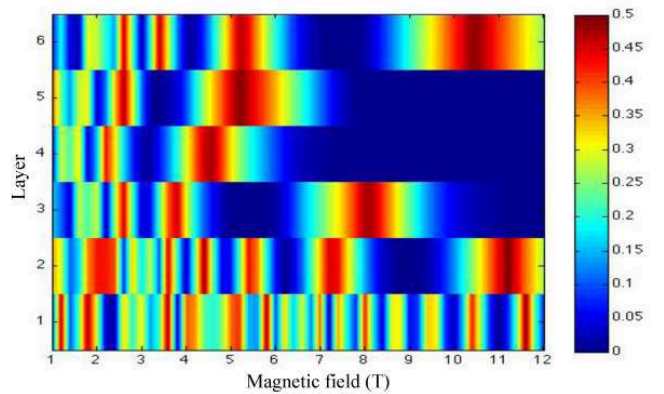}
\caption{(Color online) Filling factor modulus $\nu_i^{mod}$ as a
  function of magnetic field for $\Phi=400$ meV and $\mu=360$ meV at
  $T=30$ K, where $\nu_i^{mod}=|(\nu/4$ mod $1)-1/2|$. Note that
  $\nu_i^{mod}=0$ for $\nu_i$=$\pm 2,\pm 6, \pm 10, \cdots$, while
  $\nu_i^{mod}=1/2$ for $\nu_i$=$0,\pm 4. \pm 8, \cdots$.
The points were evaluated at zero bias voltage. 
}
\label{fig:layer_active}
\end{figure}

In addition to effects associated with exchange and correlation within
a layer, we expect that interlayer correlations play an essential role
when more than one layer is active, at least when the magnetic field
is strong and Landau levels are well developed.  
In Fig.~\ref{fig:layer_active} we plot the field dependence of the 
partial filling factor (per layer and spin) of the active Landau level:
$\nu_i^{mod}\equiv |(\nu/4$ mod $1)-1/2|$, which we refer to as the
modular filling factor.
$\nu_i^{mod}$ is defined so that it vanishes when 
a Landau level is
either completely filled or completely empty, namely at total
filling factors $\nu_i$=$\pm 2,\pm 6, \pm 10,\cdots$.  
Note that the layer is inactive when $\nu_i^{mod}=0$, and most active when
$\nu_i^{mod} = 0.5$.  Around $B=11$ T, we see that the top (6th) layer
and the 2nd layer have filling factors close to half-filled filling
factors, $\nu_6=4$ and $\nu_2=8$, respectively (in this field range
the intervening layers all have total Landau level filling factor
$\nu=2$).  This is precisely the field range in which a gap appears to
open in the top layer $N=1$ tunneling density of states in STM
studies.\cite{Joe2}  The gap could therefore be due to correlations
between $N=1$ electrons in the top layer and $N=2$ electrons in layer
2.  The participation of electrons in another layer could explain the
appearance of a gap at a partial filling factor which is not known to
support large gaps in a single-layer system.  One possible state that
is consistent with experiments is one in which coherence is
spontaneously\cite{jpenature} established between layers $2$ and $6$.
Around $B=5$ T and below, the top layer and 5th layers become active,
but due to the smaller field magnitude, strong correlation effects are
more easily suppressed by disorder.

The analysis presented in this paper highlights both advantages and
disadvantages of few layer graphene systems for physics studies.
Because the electronic degrees of freedom in all layers can play an
active role, particularly at strong magnetic fields, the physics is
extremely rich.  On the other hand, the same property makes it more
challenging to uniquely interpret observations using surface physics probes, 
like STM, which are directly sensitive mainly to top layer properties.

\acknowledgements
The work has been supported in part by the NIST-CNST/UMD-NanoCenter 
Cooperative Agreement. 
AHM was supported by Welch Foundation Grant No. F1473 and by the NSF-NRI
SWAN program.


\begin{thebibliography}{20}
\bibitem{note_errors}
Experimental peak energy positions were determined by fitting a Lorentzian 
function to the measured Landau level peaks.  The positions are averaged over 
typically 50 spectra with one standard deviation statistical uncertainties 
typically less than 1 mV, which is smaller than the plotted symbol size.
\bibitem{note1}
Note however that the temperature we use here to simulate disorder
broadening effects is much larger than measured Landau level
widths. We were unable to obtain convergence at substantially lower
temperatures. In some cases, a lack of convergence in Hartree-level 
calculations may signal that interlayer correlations play an essential 
role in the physics.
\end{thebibliography}
\end{document}